\documentclass[preprint,english,showpacs,11pt,floatfix]{revtex4}

\usepackage{babel,amsmath,amssymb,dcolumn}
\usepackage[dvips]{graphics}
\usepackage{hyperref}

\usepackage{amsfonts}
\usepackage{amssymb}
\usepackage[dvips]{graphicx}
\usepackage{latexsym}

\usepackage{dcolumn}
\usepackage{bm}

\newcommand{\be}{\begin{equation}}
\newcommand{\ee}{\end{equation}}

\begin{document}

\title{Thermodynamics of an accelerated expanding universe}

\author{Bin Wang}
\email{wangb@fudan.edu.cn} \affiliation{Department of Physics,
Fudan University, Shanghai 200433, People's Republic of China }

\author{Yungui Gong}
\email{gongyg@cqupt.edu.cn}
\affiliation{Institute of Applied Physics and College of Electronic
Engineering,
Chongqing University of Posts and Telecommunications, Chongqing 400065,
China}

\author{Elcio Abdalla}
\email{eabdalla@fma.if.usp.br} \affiliation{Instituto de Fisica,
Universidade de Sao Paulo, C.P.66.318, CEP 05315-970, Sao Paulo,
Brazil}

\begin{abstract}
We investigate the laws of thermodynamics in an accelerating universe
driven by dark energy with a time-dependent equation of state.
In the case we consider that the physically relevant part of the Universe
is that envelopped by the dynamical apparent horizon, we have shown
that both the first law and second law of thermodynamics are
satisfied. On the other hand, if the boundary of the Universe is
considered to be the cosmological event horizon the thermodynamical
description based on the definitions of boundary entropy and temperature
breaks down. No parameter redefinition can rescue the
thermodynamics laws from such a fate, rendering the cosmological event
horizon unphysical from the point of view of the laws of thermodynamics.
\end{abstract}


\maketitle

\section{Introduction}

Numerous and complementary cosmological observations tell us that our
universe is experiencing today an accelerated expansion \cite{1}. From the
Wilkinson Microwave Anisotropic Probe (WMAP) results, such an acceleration
is driven by a so-called dark energy (DE) \cite{1-2}. Such a new element of the
universe, capable of accelerating, must, in accordance with the Friedman
equation (unless we adopt wide modifications of Einstein gravity), have a
pressure less than minus one third of the energy density. In view
of the large reinterpretation of concepts implied by this energy it has
been even baptised under the suggesting name of quintessence. 

The DE has been sought within a wide range of physical phenomena, including a 
cosmological constant whose equation of state is $P/\rho\equiv w_D=-1$, 
the above mentioned quintessence with $-1<w_D<-1/3$,
or an exotic field with $w_D<-1$ \cite{2}, which can lead to even more
strange and undesirable effects such as a {\it Big Rip}, a future unavoidable
singularity of space-time. An equation of state which
gradually changes with the cosmological time is also a realistic
possibility to explain the acceleration \cite{9,10,we}. Except for the so
far well known fact that DE has a negative pressure, the nature of this
energy component still remains a complete mystery.

In this new conceptual set up, one of the important questions concerns the
thermodynamical behaviour of an accelerated expanding universe driven by a
DE. Some recent discussions on this topic can be found in \cite{3}-\cite{6}. 
With the cosmological constant, the universe is an ethernally accelerating
de Sitter (dS) space. 

On the other hand, defining a quantum theory for General Relativity turns
out to be a problem closely related to the connection between General
Relativity itself and thermodynamics. The first such hint has been obtained 
long ago by Bekenstein \cite{bekenstein} who outlined the laws of 
thermodynamics in the presence of Black Holes which turned out to be an
equivalent to the laws of Black Holes mechanics \cite{hawking}. Einstein 
Equation can even be derived from the proportionality of entropy and 
horizon area (among some further technical details) \cite{jacobson}. 
In recent years, Black Holes entropy was used as a means to a new 
reformulation of gravity in terms of the holographic principle 
\cite{thooftetal}. Relations between the entropy in the so called gravity 
bulk and a boundary Conformal Field Theory have been derived as a 
consequence of such a relation \cite{verlindeetal}, promoting such ideas 
as the most influencial in recent years' Quantum Field Theory.

The thermodynamics corresponding to an accelerated de Sitter
Universe was studied some years ago \cite{gibbons}. There is a
cosmological event horizon, analogous to a black hole horizon, which can
be associated with thermodynamical variables. Supposing that some energy
passes through the cosmological event horizon, the definitions of Black
Hole temperature and entropy imply that the first law of thermodynamics is
valid. Following earlier work, quantum gravity in de Sitter space has been
related to a Conformal Field Theory on the space-like boundary of de
Sitter space \cite{strominger,abdetal}. It has been argued that the
bulk/boundary relation might be a consequence of general aspects of
Quantum Field Theory \cite{schroer}.

For the accelerating universe driven by quintessence with constant
pressure to energy densities $w_D$ in the range $-1<w_D<-1/3$, which is
called Q-space in \cite{3}, Bousso argued that a thermodynamical
description is approximately valid and he has shown shown that the first
law of thermodynamics holds at the apparent horizon of the
Q-space. The first law has also been derived on the apparent horizon of
FRW universe with spatial curvatures \cite{caikim}. There is a subtlety between
the definitions of the apparent horizon and of the event horizon of the 
universe. In the dS universe they degenerate, while they are separated 
entities in quasi-dS universes. The question one may raise is whether the 
first law holds for both, the cosmological event horizon and the apparent 
horizon, with the definitions of temperature and entropy that we find in
the de Sitter horizon. Further, when the DE equation of state is 
time-dependent, especially with the transition from $w_D>-1$ to $w_D<-1$ 
as indicated by recent observation \cite{9}, it would be interesting 
to investigate the first law of thermodynamics, or the entropy and
temperature definitions at cosmological horizons.

The study of the first law of thermodynamics in the accelerated expanding 
universe serves as the first motivation of the present paper. Besides we 
are also going to study the expression of the entropy enveloped by 
cosmological horizons and examine the second law of thermodynamics derived
from such a definition for an accelerating universe. Our 
attention will be focused on the accelerating universe driven by DE with 
the transition equation of state described in the recent model \cite{10}.

\section{The First Law of Thermodynamics}

In this section we investigate the basic thermodynamic properties of 
accelerating universes. such as entropy, energy and temperature. We 
examine the first law of thermodynamics on
the apparent horizon as well as
on the cosmological event horizon. We begin by studying the thermodynamics
of the Q-space with constant
equation of state for DE, that is, $-1<w_D<-1/3$.

\subsection{Q-space with constant equation  of state for the DE 
($-1<w_D<-1/3$). }

The dynamical evolution of the scale factor and the matter density is
determined by the Einstein equations, which can be written in the form
\be           
(\dot{a}/a)^2=8\pi\rho/3, \hspace{1cm}\ddot{a}/a=-4\pi(\rho+3P)/3,
\ee
where the DE energy density $\rho$ and the pressure $P$ obey
$P=w_D\rho$.

Defining $\epsilon=3(w_D+1)/2$, for a constant equation of state
we have $a(t)=t^{1/\epsilon}$ and $ \rho(t)=3/(8\pi \epsilon^2 t^2)$, where
$0<\epsilon<1\; (-1<w_D<-1/3)$. This set describes the accelerating 
Q-space \cite{3}, For $\epsilon>1$ we the decelerating universe while 
$\epsilon=0$ corresponds to the de Sitter space.

The event horizon for the Q-space is
$R_E=a\int_t^{\infty}dt/a=-\epsilon t/(\epsilon-1)$. The apparent horizon
reads $R_A=1/H=\epsilon t$. The horizons do not differ much, they relate 
by $R_A/R_E=1-\epsilon$. Except for the phantom universe, $w< -1$, which 
we have to avoid for a constant equation of state in view of the {\it Big
  Rip} (we actually avoid this case here), the apparent horizon is 
smaller than the event horizon by a fixed ratio close to unity.
Neither the event horizon nor the apparent horizon changes significantly
over one Hubble time,
$t_H\dot{R_X}/R_X=\epsilon <1, X=A,E$. It is thus possible to
use the equilibrium thermodynamical theory here \cite{3}, especially for a
(very) accelerated phase. We shall take, as in the black hole case,
\be   \label{2}
S_X=\pi R_X^2, \hspace{1cm} T_X=1/(2\pi R_X) \hspace{1cm} X=A, E,
\ee
to describe the temperature of the horizon and the entropy.

Now let us examine the first law of thermodynamics on these two horizons.
For the apparent horizon
the first law was obtained in \cite{3}. The amount of energy crossing the
apparent horizon during the time interval $dt$ is
\be
-dE=4\pi R_A^2 T_{ab}k^a k^b dt=4\pi R_A^2\rho(1+w)dt=\epsilon dt\quad
.\label{3} 
\ee
The apparent horizon entropy increases by the amount
\be        
dS_A=(2\pi R_A)\dot{R_A}dt=(2\pi R_A)\epsilon dt\quad .\label{4}
\ee
Comparing (\ref{3}) with (\ref{4}) and using the definition of the temperature, 
the first law on the apparent horizon, $-dE=T_A dS_A$, was confirmed.

We now work with the cosmological event horizon. The total energy
flow through the event horizon can be similarly got as
\be            \label{5}
-dE=4\pi R_E^2\rho(1+w)dt=\epsilon dt/(1-\epsilon)^2.
\ee
The entropy of the event horizon increases by
\be           \label{6}
dS_E=2\pi R_E \dot{R}_E dt=\frac{2\pi R_A}{(1-\epsilon)^2}\epsilon dt,
\ee
and using the Hawking temperature for the event horizon we obtain
\be                   \label{7}
T_E dS_E=\epsilon dt/(1-\epsilon).
\ee
We do not obtain the first law with the above definitions 
since the energy income is larger by a factor  $1/(1-\epsilon)$.

One may argue  that the temperature of the bath should be the
same using the event and apparent horizon, which should be the local
effect on the observer. If we choose $T_A$, the temperature on the
apparent horizon, which is larger exactly by a factor $1/(1-\epsilon)$, 
the difference $1-\epsilon$ disappears and the first law
$-dE=TdS$ looks correct in this form.

But the problem is not solved. If we extend the discussion to the
time-dependent equation of state DE, even if we use the local temperature 
on the apparent horizon, we shall see that the first
law of thermodynamics still cannot be rescued on the event horizon, with
the usual definitions of the temperature and entropy.

\subsection{Accelerating universes driven by DE with time-dependent
  equation of state.} 

We shall use the holographic DE model \cite{11}. The generalization of this
model by considering interaction between DE and Dark matter (DM) has
recently been discussed in \cite{10}. For simplicity, we neglect the
interaction between DE and DM for the moment, but the result below also
holds if we include the interaction. 

The event horizon in this case is given by the expression
$R_E=a\int_a^{\infty} da/(Ha^2)=c/(\sqrt{\Omega_D}H)$. The last equality
was gotten by considering the holographic DE $\rho_D=\Omega_D 3H^2=3c^2
R_E^{-2}$ \cite{11}. The apparent horizon is $R_A=1/H$. The relation
between apparent and event horizon is $R_A/R_E=\sqrt{\Omega_D}/c$ and the
apparent horizon is in general smaller than the event horizon. If $c=1$ and
$\Omega_D=1$  both horizons are the same, $ R_E=R_A$. 

Neglecting the interation between the DE and DM, the evolution of the DE
was obtained as \cite{11}
\be    \label{8}
\Omega_D'=\Omega_D^2(1-\Omega_D)\lbrack \frac1{\Omega_D}+
\frac2{c\sqrt{\Omega_D}}\rbrack\quad ,
\ee
and the equation of state of the DE has the form
\be          
w_D=-1/3-2\sqrt{\Omega_D}/(3c).
\ee
With the evolution of the DE, $w_D$ changes with cosmological time.

We now examine how much the horizon will change over one Hubble time. For
the apparent horizon
\be          \label{10}
t_H\frac{\dot{R_A}}{R_A}=aH\frac{d(H^{-1})}{da}=H\frac{dH^{-1}}{dx},
\ee
where $x=\ln a$. From the Friedmann equation, we have
$1-\Omega_D=\Omega_m=\rho_m/(3H^2)=\Omega_{m0}H^2_0 H^{-2}a^{-3}$, therefore
$H^{-1}=a^{3/2}\sqrt{1-\Omega_D}/(H_0\sqrt{\Omega_{m0}})$.
Eq(\ref{10}) can be rewritten as
\be             
t_H\frac{\dot{R_A}}{R_A}=3/2-\Omega_D'/(2(1-\Omega_D)).
\ee
Considering the evolution of the DE, Eq.(\ref{8}), we have
\be               
t_H\frac{\dot{R_A}}{R_A}=3/2-\Omega_D/2-\Omega_D^{3/2}/c\quad .
\ee

We now calculate the change of the event horizon, which can be obtained
from 
\be                 
t_H\frac{\dot{R_E}}{R_E}=\frac{3}{2}-\frac{\Omega_D'}
{2\Omega_D(1-\Omega_D)}=1-\frac{\sqrt{\Omega_D}}{c}\quad.
\ee

We learn neither the apparent horizon nor the event horizon change 
significantly over one hubble scale. This can be seen from Fig. 1. 
Actually, the event horizon changes less than the Hubble horizon. The
equilibrium thermodynamics still can be applied here. As originally
suggested for black holes, the temperature and entropy on the horizon are
described in Eq. (\ref{2}).

\begin{figure}[hbt]
\begin{center}
\includegraphics[width=10cm]{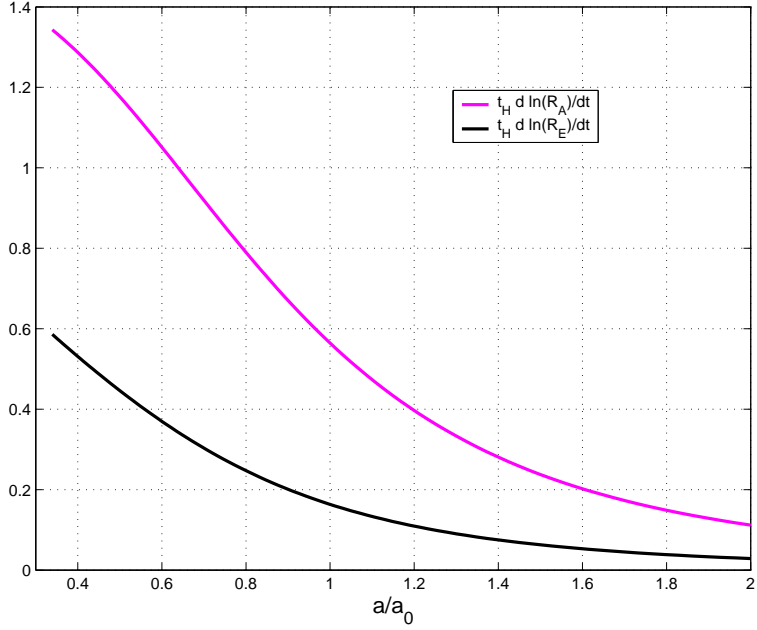}
\end{center}
\caption{The rate changes of the apparent horizon and the event horizon
  over one Hubble scale with $\Omega_{D0}=0.7$ and $c=1$.}  
\end{figure}

Now we start to investigate the first law of thermodynamics on different
horizons.
For the apparent horizon, the total amount of energy crossing the apparent
horizon during $dt$ is
\be             
-dE=4\pi R_A^2\rho(1+w)dt=3(1+w)/2 dt.
\ee
Employing
$w=\frac{P_D}{\rho}=w_D\rho_D/(3H^2)=w_D\Omega_D
=(-1/3-2\sqrt{\Omega_D}/(3c))\Omega_D$, we get
\be                      
-dE=(3/2-\Omega_D/2-\Omega_D^{3/2}/c)dt.
\ee

On the other hand, using the definitions of temperature on the horizon and
the holographic entropy we get
\be                
T_A dS_A=\dot{R_A} dt=H\frac{dH^{-1}}{dx}
dt=[3/2-\Omega_D/2-\Omega_D^{3/2}/c]dt.
\ee
Thus on the apparent horizon, the first law, $-dE=T_A dS_A$, is confirmed,
with the above definitions.

We now extend the discussion to the event horizon. The energy flow through
the event horizon reads
\be     
-dE=4\pi R_E^2\rho (1+w)dt=\frac{3c^2(1+w)}{2\Omega_D}
dt=\frac{c^2(3/2-\Omega_D/2-\Omega^{3/2}/c)}{\Omega_D}dt.
\ee
To examine the first law, let us compute $T_EdS_E=\dot{R_E} dt$. Using
\be                
\dot{R_E}=\frac{d}{dt}(\frac{c}{\sqrt{\Omega_D}H})
=H\frac{d}{dx}(\frac{c}{\sqrt{\Omega_D}H})=\frac{3c}
{2\sqrt{\Omega_D}}-\frac{c\Omega_D'}{2\Omega_D^{3/2}(1-\Omega_D)}
\ee
and (\ref{8}), we have $\dot{R_E}=c/\sqrt{\Omega_D}-1$.

We thus see that the first law cannot be satisfied in this form.  It
cannot be rescued even
by using the same temperature of the bath on the apparent horizon,
$T=1/(2\pi R_A)$, as we could have done in the Q-space, since
\be           
T_AdS_E=\frac{R_E}{R_A}\dot{R_E}dt=\frac{c} {\sqrt{\Omega_D}}
\dot{R_E}dt=[\frac{c^2}{\Omega_D}-\frac{c}{\sqrt{\Omega_D}}]dt\quad .
\ee
Therefore, the first law of thermodynamics cannot hold at the event
horizon with the usual definition of entropy and temperature!!

Although the dynamical difference between the apparent horizon and the
event horizon is not large, the difference of their thermodynamical 
properties really is. In the de Sitter universe, this problem was hidden,
since in that case the event horizon and the apparent horizon were
degenerate. This problem becomes especially sharp with a dynamical
equation of state. One reason could be that the
first law may only apply to variations between nearby states of local
thermodynamic equilibrium, while the event horizon reflects the global
properties of spacetimes. At the event horizon, maybe the temperature and
the entropy are ill-defined. In the nonstatic spacetime, the horizon
thermodynamics are not simple, the notion of the surface gravity
maybe ill-defined as argued in the study of the quasi-dS geometry in the
inflationary universe \cite{12} and the definition of temperature and
entropy by means of the relations found by Bekenstein and Hawking are
possibly wrong.

\section{Entropy and the Second law of thermodynamics}

We will study the entropy enveloped by cosmological horizons. To exhibit
the entropy of the accelerating universe with DE equation of state
$w_D<-1$ together with $w_D>-1$ cases, we use our interacting holographic
DE model \cite{10}. 

The entropy of the universe inside the horizon can be related to its
energy and pressure in the horizon by Gibb's equation \cite{6}
\be          
TdS=dE+PdV.
\ee
For the apparent horizon, considering $V=4\pi R_A^3/3, E=4\pi \rho
R_A^3/3=R_A/2, P=w\rho=w_D\Omega_D 3H^2/(8\pi)$, we have
\be               
TdS=dR_A/2+(3/2) w_D\Omega_D dR_A.
\ee
Using $T=1/(2\pi R_A)$, which is the only temperature scale we have at our
disposal, we get
\be            
dS=\pi(1+3w_D\Omega_D)R_A dR_A.
\ee
Thus we see that the entropy enveloped by the apparent horizon is $S\sim
R_A^2$.

Now we need to solve the equation to see the evolution of the entropy.
Since
$R_AdR_A=R_A\frac{dR_A}{dx}dx=-\frac{1}{H^3}\frac{dH}{dx}dx$, we have
\be             \label{23}
\frac{dS}{dx}=-\pi(1+3w_D\Omega_D)H^{-3}\frac{dH}{dx}=
-\pi(1-3b^2-\Omega_D-2\Omega_D^{3/2}/c)H^{-3}\frac{dH}{dx}=-2\pi
q H^{-3}\frac{dH}{dx}
\ee
where we have used the evolution of the DE in our interacting holographic
DE model \cite{10}
\be         
\frac{\Omega_D'}{\Omega_D}=1-3b^2-\Omega_D-
\frac{2\Omega_D^{3/2}}{c}+\frac{2\sqrt{\Omega_D}}{c}
\ee
and the equation of state of the DE
\be             
w_D=-\frac{\Omega_D'}{3\Omega_D(1-\Omega_D)}-\frac{b^2}
{\Omega_D(1-\Omega_D)}=-\frac{1}{3}-\frac{2}{3}
\frac{\sqrt{\Omega_D}}{c}-\frac{b^2}{\Omega_D}\quad .
\ee
Above, $b^2$ is the coupling between DE and DM. The evolution of the Hubble
parameter in our model was described as
\be        
\frac{d\ln
H}{dx}=-3/2+3b^2/2+\Omega_D/2+\Omega_D^{3/2}/c=
-(3-3b^2-\Omega_D-2\Omega_D^{3/2}/c)/2\quad .
\ee
In (\ref{23}) $q$ is the deceleration parameter, given by
\be        
q=-\dot{H}/H^2-1=-3[-1-w_D-(1-\Omega_D)/\Omega_D]
\Omega_D/2-1=(1-3b^2-\Omega_D-2\Omega_D^{3/2}/c)/2\quad .
\ee

To accommodate the transition of the DE equation of state from $w_D>-1$ 
to $w_D<-1$ at
recent stage as indicated by observations \cite{9},
we constrained our model parameters $c$ from holography in the range
$\sqrt{\Omega_D}<c<1.255$ and $b^2$
of the coupling between DE and DM within the range
$1.4(1-\sqrt{0.7}/c)/3<b^2<8c^2/81$ \cite{10}. Within these parameter
space, we have
$\Omega_D'>0$ and $H'<0$. At early stage, $\Omega_D$ is small and the
universe was in the deceleration era with $q>0$, so we obtain $dS/dx>0$.
However at late time, DE starts to dominate, the universe evolved in an
accelerated expansion with $q<0$, thus we get $dS/dx<0$.
Therefore the entropy of the universe enveloped by the apparent horizon
increased first and then started to decrease.
The entropy will decrease to a  negative value  at a late era of the
universe when it is dominated by phantom fields,
which is consistent with those observed in \cite{5}\cite{6}. The behavior
of this entropy is shown in the dotted line in Fig 2.

\begin{figure}[hbt]
\begin{center}
\includegraphics[width=10cm]{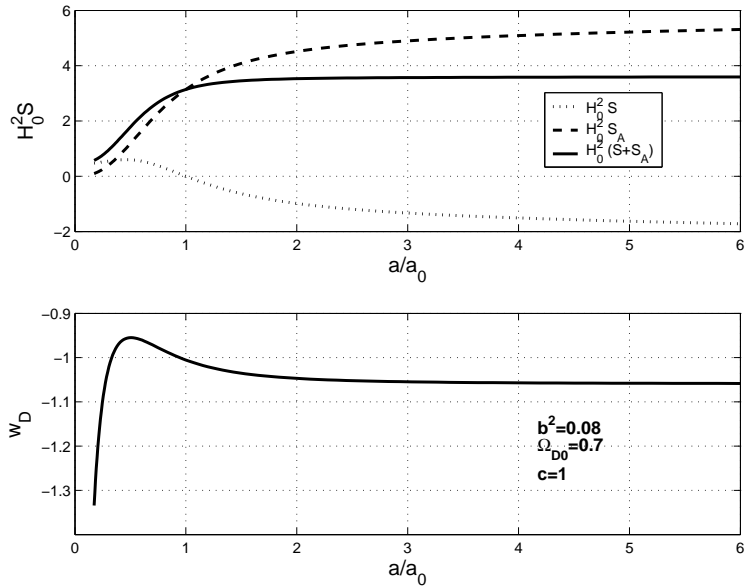}
\end{center}
\caption{The evolution of entropies and $\omega_D$ with the parameters
  $b^2=0.08$ and $c=1$ and the initial conditions $\Omega_{D0}=0.7$ and
  $H^2 S_0=10^{-30}$.} 
\end{figure}

In addition to the entropy in the universe, there is a horizon entropy of
the apparent horizon $S_A=\pi H^{-2}$. The evolution of this
horizon entropy is shown in the dashed line in Fig.2. At the early stage
we see that $S>S_A$, which is in violation of the
holographic principle. This tells us that although our interacting
holographic DE model \cite{10} is sucessful to describe the
late time accelerating universe with the transition equation of state of
DE, this model is not appropriate to describe the early
deceleration of the universe.

In order to check the generalized second law (GSL) of gravitational
thermodynamics, we have also plotted the evoluation of $S+S_A$,
the solid line in Fig.2. It is easy to see that entropy of matter and
fluids inside the apparent horizon plus the entropy of the apparent 
horizon do not decrease with time. Thus
the GSL is respected in the universe enveloped by the apparent horizon.

It is of interest to examine the GSL in the universe enveloped by the
cosmological event horizon. The evolution of the entropy
inside the event horizon can be gotten similarly as that of the apparent
horizon
\be                  
\frac{dS}{dx}=-\pi
c^4(1+3w_D\Omega_D)H^{-3}\Omega_D^{-2}\frac{dH}{dx}- \frac{3\pi}{2}c^4
(1+w_D\Omega_D)H^{-2}\Omega_D^{-3}\frac{d\Omega_D}{dx}\quad ,
\ee
while the evolution of the geometric entropy on the event horizon reads
\be                     
\frac{dS_E}{dx}=-2\pi c^2 H^{-3}\Omega_D^{-1}\frac{dH}{dx}-\pi c^2
H^{-2}\Omega_D^{-2}\frac{d\Omega_D}{dx}\quad .
\ee
We thus find
\be         
\frac{d(S+S_E)}{dx}=\frac{\pi}{2}c^4H^{-2}\Omega_D^{-2}[AB-CD],
\ee
where $A=2q+2\Omega_D/c^2$, $B=2+2q$, $C=2q+2\sqrt{\Omega_D}/c$,
$D=2+2q+2\Omega_D/c^2$. Since in our model it is required that
$\sqrt{\Omega_D}<c$, we have $A<C$ and $B<D$. Thus, for the universe
enveloped by the cosmological event horizon we have $d(S+S_E)/dx<0$ and the
GSL breaks down, with the usual definitions of entropy for the boundary.

\section{Conclusions and discussions}

The apparent horizon of the universe always exists and the thermodynamical
properties related to the apparent horizon has been studied by several 
authors, including in a quasi-dS geometry of inflationary universe
\cite{12} and a late time accelerating Q-space \cite{3}.
In this paper we have extended the investigation to an accelerated
expanding universe driven by DE of time-dependent equation of state.
We have confirmed that the first law of thermodynamics still holds for
this dynamic apparent horizon. Besides we have also examined the GSL
of gravitational thermodynamics of the accelerating universe driven by the
interacting holographic DE \cite{10} with the transition equation of state.
The entropy of matter together with fluids inside the apparent horizon
plus the entropy of the apparent horizon always increase with time,
which shows that the GSL is protected in the accelerating universe
enveloped by the apparent horizon.

In the usual standard big bang model a cosmological event horizon is
absent. The thermodynamics related to the cosmological
event horizon was studied in the dS space \cite{gibbons}, however in dS space,
the cosmological event horizon and apparent horizon degenerate.
In a general accelerating universe driven by DE with equation of state
$w_D\neq -1$, the 
cosmological event horizon separates from that of the
apparent horizon. In this paper we have tried to apply the usual
definition of the temperature and entropy as that of the apparent horizon
to the cosmological event horizon and examine the first and the second laws
of thermodynamics. In contrast to the case of the aparent horizon,
we are surprised to find that both the first and second law of
thermodynamics break down if we consider the universe to be enveloped by 
the event horizon with the usual definitions of entropy and temperature.
The break down of the first law could be attributed to the possibility
that the first law may only apply to variations between nearby
states of local thermodynamic equilibrium, while the event horizon
reflects the global spacetime properties.
Besides in the dynamic spacetime, the horizon thermodynamics is not as
simple as that of the static spacetime. The event horizon and
apparent horizon are in general different surfaces. The definition of
thermodynamical quantities on the cosmological
event horizon in the nonstatic universe are probably ill-defined. This was
first argued in the quasi-dS geometry of the
inflationary spacetime \cite{12}.

Furthermore we would like to point out that the apparent horizon serving as 
a specific surface was observed by Bousso in studying the covariant 
entropy bound \cite{bousso99}. The apparent horizon was singled out as the
largest surface whose interior can be treated as a Bekenstein system,
which satisfies the Bekenstein's entropy/mass bound $S\leq 2\pi RE$ and
Bekenstein's entropy/area bound $S\leq A/4$. In the region surrounded 
by the surface outside the apparent horizon, one could satisfy Bekenstein
entropy/mass bound but break Bekenstein entropy/area bound, which
indicates a breakdown of the treatment of the enclosed region as a
Bekenstein system. Since Bekenstein bounds are universal, all
gravitational stable spacial regions with weak self-gravity should 
satisfy Bekenstein bounds and the corresponding thermodynamical system is a
Bekenstein system. This should also be true in cosmological situations.
For the dS situation, since the cosmological event horizon coincides 
with the apparent horizon, the region enclosed by the cosmological 
horizon satisfies Bekenstein bounds. However for the accelerating universe 
driven by DE with $w_D\neq -1$, from our equations (\ref 5) and (\ref 7)
above, for example, we see that at the event horizon $dE>dR_E$, or
$E>R_E$. Although the Bekenstein entropy/mass
bound can be satisfied, the Bekenstein entropy/area bound is violated,
since $S=2\pi ER_E>\pi R_E^2$. Thus the thermodynamic system outside 
the apparent horizon is no longer a Bekenstein system and the
usual thermodynamic description breaks down.

To our knowledge, this thermodynamical problem related to the event 
horizon in the accelerating universe driven by the DE has not been 
found before. The solution of this problem requires further knowledge,  
which we do not have at disposal at this moment. 

\begin{acknowledgments}
This work was partially supported by  NNSF of China, Ministry of
Education of China, Ministry of Science and Technology of China
under grant No. NKBRSFG19990754 and Shanghai Education Commission.
Y. Gong's work was supported by NNSFC under grant No. 10447008,
CSTC under grant No. 2004BB8601, CQUPT under grant No. A2004-05
and SRF for ROCS, State Education Ministry. E. Abdalla's work was
partially supported by FAPESP and CNPQ, Brazil. E.A. wishes to thank
G. Matsas, S. Salinas and W. Wreszinski for discussions.
\end{acknowledgments}


\end{document}